# QUANTUM SIMULATION & OPTIMIZATION OF WATER DISTRIBUTION NETWORKS

Carlos ROMERO ROCHA[1,2], Nicolas RENAUD[1,2], Koen LEIJNSE[1], Samuel VAN BEEK[1], Mario CASTRO-GAMA[3]

[1] Quantum Application Lab
email: s.vanbeek@uva.nl
email: k.leijnse@uva.nl

[2] Netherlands eScience Center
email: n.renaud@esiencecenter.nl
email: carlos.romerorocha@tno.nl[*]

[3] Vitens
email: mario.castrogama@vitens.nl

## ABSTRACT

To compute models for Water Distribution Networks (WDN), a large system of non-linear equations needs to be solved. The hallmark algorithm for computing these models is the Newton-Raphson Global Gradient Algorithm (NR-GGA), which solves these systems iteratively. Even so, large networks can take multiple days to model and the complexity of networks is only expected to increase in the future. It is therefore important to explore different algorithms using innovative technologies, to improve the tractability of modelling large networks. Quantum computing is such an innovative technology that is still in its early stages of development, paired with a different computational paradigm. In this research we have determined the feasibility of using quantum computing algorithms as a subroutine of NR-GGA and alternatively for replacing NR-GGA with a quantum algorithm in its entirety. Calculations were run on emulators of gate-based quantum computers and using simulated annealing while models were tested on small 2-loop networks of 7 nodes. To improve NR-GGA, three different quantum subroutines were used: the hybrid Variational Quantum Linear Solver (VQLS) showed the best results on this small dataset. To replace NR-GGA by a quantum algorithm in its entirety, we also used a Quantum Annealing-based approach for simulating higher-order polynomials to determine the Least-Cost Design (LCD) of a small 3 node network.

**Keywords:** *Least-Cost Design; Newton-Raphson Global-Gradient-Algorithm; Quantum Computing; Water Distribution Network Modelling*

## 1. Introduction

Large water distribution system operators are faced with network models that are intractable to compute in their full scale in short time periods and must instead compute smaller networks over larger time periods. These models are important for understanding e.g., pipe friction parameters, determining least-cost designs, finding leaks, developing digital twins and more (Sonaje&Joshi, 2015).

The primary focus of this project has been to develop and test quantum algorithms that could help optimize water distribution by modelling the water flowrate and pressure of Water Distribution Networks (WDN's). Traditional methods may struggle with the complexity of predicting and managing water flow, especially under varying loading conditions and for large networks. The use of quantum computing may improve decision-making processes and enhance the development of water distribution resources (Dalzell et al., 2023).

Although large-scale fault-tolerant quantum computing is still several years off, an advantage of using quantum computers for niche use-cases may be found in the near future (Groenland, 2024). Quantum computers introduce with them a new computational paradigm: hardware-use, software stacks, network communication, use-cases and security all need to be re-thought and understanding and implementing these changes may take many years for end-users of quantum technology. As a result, it is relevant to start working on these challenges through early Proof-of-Concept (PoC) development of quantum computing applications. These PoC's tend to focus on determining *feasibility* of solving a certain computational challenge on a quantum computer and while there is in some cases theoretical evidence for asymptotic speed-ups, claims regarding scalability for practical use-cases are harder to make. This difficulty arises in part from the use of heuristic algorithms, either quantum or classical, and in another part from the continuous advancement of state-of-the-art classical solutions that any potential quantum solution will have to beat in some metric, like speed or accuracy.

---
[*] Contributions/Work done while affiliated with Netherland eScience Center

This report summarizes the research conducted, the methods used, and the results obtained. It provides an overview of the challenges encountered and the potential benefits of applying quantum computing to water management. The findings offer a practical perspective on the feasibility of integrating this emerging technology into real-world applications for simulating WDN's. The paper concludes with an outlook for the scalability and feasibility of using quantum computing for modelling WDN's as the technology develops.

## 2. Methodology

### 2.1 Quantum Computing

Quantum computing is a new paradigm of computing that relies on principles of quantum mechanics, such as superposition and entanglement, to solve certain computational tasks in less computational steps than a classical computer could. These same quantum principles that lend quantum computers additional computational strengths, also make the development of the technology an extremely complex task. New hardware, known as the quantum computer, is required where quantum bits of information (qubits) are represented by quantum systems, such as individual photons or atoms, that are hard to manipulate and prone to error. On top of that, different computational models have been developed to describe the quantum algorithms that may be executed on these quantum computers and a new quantum software stack is required to write quantum algorithm instructions to the quantum computer.

There are many different approaches to the development of quantum computing hardware. Most cutting-edge realizations of the quantum computer have in the hundreds of error-prone qubits, while some large-scale quantum applications, such as breaking RSA-encryption efficiently will require in the tens of thousands stable, error-resistant qubits (Gouzien&Sangouard, 2021). The technology is still in early development and understanding the nature of problems that can benefit from this technology proves challenging. Typical candidates for future quantum speed-ups over classical state-of-the-art algorithms are problems involving e.g., combinatorial optimization, simulating quantum effects and solving large linear systems.

In this experimental stage of quantum computing development, alternative approaches towards developing quantum machines have emerged. These technologies sacrifice the ability to execute any possible quantum algorithm, i.e. they are non-universal, but focus on executing a particular class of quantum algorithms. The benefit is that these technologies are often at a higher Technology Readiness Level (TRL) than their universal counterparts. The most well-known and relevant for this project is the quantum annealer, a non-universal quantum computer that excels at solving combinatorial optimization problems through Quantum Annealing (QA) (Yarkoni et al., 2022).

### 2.2 Quantum Computing for Simulating WDN's

Two different approaches to modelling WDN's using quantum computers were explored in this work, using 4 different quantum algorithms.

The first approach uses a hybrid method, starting at the crucial Newton-Raphson Global Gradient Algorithm (NR-GGA) (Todini&Pilati, 1998; Todini&Rossman, 2012) and using one of three quantum algorithms as a subroutine. NR-GGA is an iterative algorithm that finds the solution of a large non-linear system of equations by iteratively updating the solution vector using the gradient of a loss function. Computing the gradient involves solving a large linear system of equations, at which quantum computers are expected to be proficient (Dalzell et al., 2023), and which is done using a quantum algorithm in our Quantum version of NR-GGA (QNR).

The first of the three quantum subroutines, VQLS, is a parameterized quantum algorithm that is in itself a hybrid classical-quantum algorithm whose gate-parameters are optimized within a classical optimization loop (Bravo-Prieto et al., 2023). The algorithm can be described as a quantum circuit with parameterized gates whose parameters are updated using a loss function. Quantum computers are prone to errors and the error-behavior of qubits and couplers between qubits differs between quantum devices. Learning algorithms like VQLS are useful at the current experimental stage of quantum computing for their capacity to adapt well to inter-device differences through learning.

The second quantum subroutine, HHL, is a well-known quantum algorithm that is proven to asymptotically solve linear systems faster under 'specific conditions' than any known classical method (Harrow et al., 2009): Given a quantum oracle that can efficiently query entries of the input matrix of the linear system to be solved, and provided that the input matrix is sparse and well-conditioned, HHL can output a quantum state containing the solution vector in its amplitudes in logarithmic computational steps. The fastest known classical algorithm solves linear systems in a polynomial number of steps. When not all these conditions are present a speed-up is no longer guaranteed, but still possible, and empirical evidence becomes increasingly valuable.

The third subroutine we used in QNR, QUBO-LS, is an algorithm that makes use of QA technology. As stated in the previous subsection, QA is used for solving combinatorial optimization problems and does this by finding the ground state of the Ising model for a quantum system through annealing. The ground state energy of the Ising model is given by a Quadratic Unconstrained Binary Optimization function (QUBO). Any function that can be reduced to a QUBO can then be solved using quantum annealing and this reduction exists for the problem of finding the solution of a linear system of equations. The effectiveness of using QA is often determined by how much overhead is induced in reducing a computational problem to QUBO form (Yarkoni et al., 2022).

*2.3 Quantum Annealing for Optimizing Water Networks*

In our second approach, NR-GGA is replaced entirely by a quantum algorithm also based on QA. The task of modeling WDN's is often part of other analyses of WDN's, such as optimizing network installation, calibrating friction parameters or detecting leaks. Reducing the WDN-modeling task to a QUBO form that lets us model WDN's on using QA, also provides us with the possibility to introduce additional optimization parameters such as pipe-widths in order to minimize e.g., network installation costs. To demonstrate this, we compute the Least-Cost Design (LCD) of a network (Babayen&Kapelan, 2005). The optimization problem is non-linear and non-quadratic and as a result, additional overhead is induced in translating the objective function into QUBO form. Solving a higher order polynomial optimization function using QA is based on work by Chang et al. from 2019.

Developed codes have been published on the Quantum Application Lab GitHub[1] and made open-source. For the QNR- code, EPANET (Rossman et al., 2020) and WNTR (Klise et al., 2020) were used as a basis.

3. **Results and Discussion**

Using real quantum hardware is possible, but changes the focus of the research, as it requires a lot of attention to be spend on applying error-mitigation techniques and reduces the potential problem input sizes. As such, quantum emulators were used to get a better grasp of feasibility for simulating WDN's using quantum computers.

*3.1 QNR Stability tests*

To simulate WDN's we ran a series of stability tests, computing water pressures and flowrates over varying scenarios. We varied pipe-length, pipe-diameter, roughness coefficients and demands. For each variable we ran 350-500 scenarios, sampling variables from within a specified range uniformly. Every point on the graphs of Figures 1-3 represents a calculated pressure for one of the nodes in the network after a full run of QNR with the specified quantum subroutine, with differently colored points representing models with different loading conditions. Simulations were run for networks with 7 nodes and 8 pipes and 2 loops, referred to in text as the 2-loop network. Results were benchmarked with classically computed values using EPANET.

---

[1] Github: https://github.com/QuantumApplicationLab, QNR: https://github.com/quantumapplicationlab/wntr-quantum, Higher order polynomial QA: https://github.com/QuantumApplicationLab/qubops

*VQLS*

As seen in Fig. 1, the results obtained with VQLS are in very good agreement with the reference values. VQLS appears to be able to correctly solve the linear systems obtained at each iteration of the NR-GGA algorithm. As a result, QNR-VQLS reaches the correct solution. These results are also valid in a wide range of operational conditions, showing the stability of QNR-VQLS. It is worth repeating that these simulations were performed on a quantum emulator, i.e. classical high performance computing resource capable of reproducing the behavior of quantum computers. Noise sources that characterize current generation quantum computers were not accounted for here. The deployment of similar simulations on real quantum hardware is left for future developments.

Since variational quantum algorithms have access to a more complex parameter space compared to classical machine learning algorithms of the same size, it is expected that QNR-VQLS will scale well. However, to determine a better timeline of the scalability, algorithms need to be run on real quantum hardware, not quantum emulators. Quantum hardware is error-prone and advanced error mitigation techniques are required to get the most out of understanding its limitations. Furthermore, there are still theoretical challenges that need to be overcome for variational quantum algorithms to scale up efficiently, such as finding efficient methods for backpropagating gradient computations (Abbas et al., 2023).

*HHL*

Figure 2 shows the results of our simulations using the HHL algorithm as part of the QNR-HHL in different operational conditions. As seen in this figure, the results obtained with HHL do not reproduce the results obtained with the classical linear solver. While a certain degree of correlation can be observed between the true values of the head pressure and flow rate with those obtained when using the HHL solver, the spread of the HHL values is much greater than the tolerance region of 10%.

The probable explanation for this result has to do with the condition numbers of the linear systems found along the Newton-Raphson optimization trajectory. The HHL circuit used during the simulation was parametrized to obtain a correct solution of the linear systems obtained in the first few steps of QNR-HHL at a reasonable computational cost. However, if the condition numbers of the linear systems at a later stage in QNR-HHL exceed the capabilities of this quantum circuit, a poor solution of the linear system will be obtained, preventing QNR-HHL to converge to the correct solution. Different solutions can be devised to circumvent this issue, from increasing the accuracy of the quantum circuit, and therefore the number of qubit and the compute time, from the very beginning of the NR-GGA to dynamically recreating the quantum circuit to fit the requirement of the linear system. While these solutions could be implemented in our software, they fell outside of the scope of this research. Furthermore, it is well-known that HHL algorithms require high circuit depths, making testing on real hardware, where large circuits lead to compounding errors, difficult (Harrow et al., 2009).

*QUBO-LS*

Figure 3 shows the results of our simulations using the QUBO-LS algorithm as part of the QNR-QUBO solver in different operational conditions. As seen in this figure, the results obtained with QUBO linear solver are in good agreement with the reference values. The optimization of the QUBO problem might converge towards a local minimum of the objective function. This leads to the presence of outliers. While a more careful optimization schedule should limit the occurrence of these sub-optimal solutions, it is impossible to completely rule them out.

When using real quantum hardware, annealing methods have the advantage of having access to a more advanced quantum technology with more available qubits, allowing for larger problem instances to be solved. However, a large overhead is induced in translating a linear system problem to a QUBO so as gate-based quantum computing technology matures, QA-based solutions for solving linear systems might become less interesting.

*3.2 Higher-order QA for WDN Simulation*

QA can be used to solve non-linear systems of equations like the ones governing the flowrates and pressure values of WDN. To do so, the higher order polynomial equations must be quadratized to allow for a QUBO expression. This quadratization is done by introducing additional variables and therefore further increasing the computational complexity (Chang et. al. 2019). When applied to the simulation of a WDN, this approach does not rely on the QNR algorithm as the flowrates and pressure values are directly obtained from the QA outcome.

We have used this approach to compute the flowrates and pressure values of the 2-loop network, as seen in Fig. 4. Several annealing runs are sampled, each providing a different associated energy, with lower energy being closer the optimal ground-state solution of the translated annealing problem. The solution with minimum energy, marked with blue dots, displays values within the 10% error range at the exception of one flow rate. The solutions for all the optimization paths lead to values of the pressure and flow rate well outside the 10% tolerance region. This illustrates the difficulty in classically optimizing the QUBO problem obtained from the hydraulics equations. A quick analysis of the problem shows that the QUBO energy landscape is characterized by pronounced local minima that are difficult to escape during the optimization, even when using our custom simulated annealing sampler. Using quantum hardware might allow to reach the global minimum more efficiently and may be explored in future studies.

*3.3 Higher-order QA for WDN Least Cost Design*

QA offers the possibility to optimize other parameters of a given WDN. This is done by treating the parameters as variables of the system and optimizing their values to minimize the cost function. We have expressed the LCD problem as a large QUBO problem where the pipe diameters are treated as variables together with the flowrates and pressure values. The cost function of the optimization was modified to minimize the overall cost of the WDN and constraints were introduced to ensure that the pressure values remained above a certain threshold. The size of the resulting QUBO problem forced us to limit our exploration to a small 3-node network with 2 pipes both with possible diameters of 0.25m, 0.5m and 1m.

Figure 5 shows the results of our LCD simulations on the 3-node network. The QUBO formulation of the LCD problem is an optimization problem and its cost function defines an energy landscape. The annealer samples solutions heuristically, ideally sampling low energy solutions, at- or close to the ground-state. Since there are only 9 combinations of pipe-widths, it is easy to determine the least-cost solution of the model, which in this case was 2 pipes of 0.5m.

As seen in this figure, most of the optimization runs converge toward the correct composition of the pipe diameters. It is important to note that, even with this composition, a large variance of the final energy is observed. This is due to the many possible local minima obtained for values of the pressure and flow that are not exact solutions of the hydraulics equation. It is also worth noting that a significant fraction of the solutions converges towards values of the pipe diameters that are not optimal. Here again the presence of local minima in the energy function is to blame.

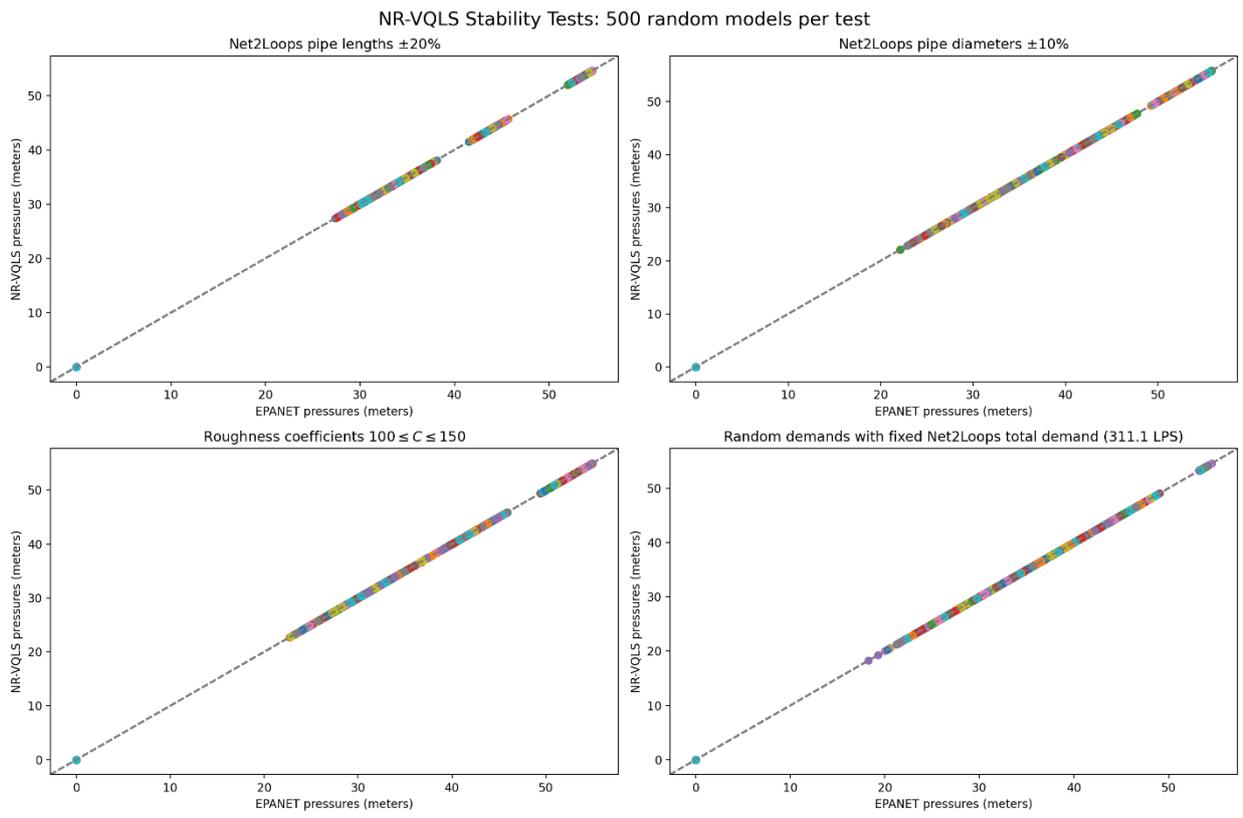

**Fig. 1.** Simulated water pressures using QNR-VQLS for a 2-loop network for a wide range of pipe diameters, length, roughness and demand. The x-axis marks the reference values obtained with EPANET while the y-axis shows the values obtained with the QNR-VQLS approach.

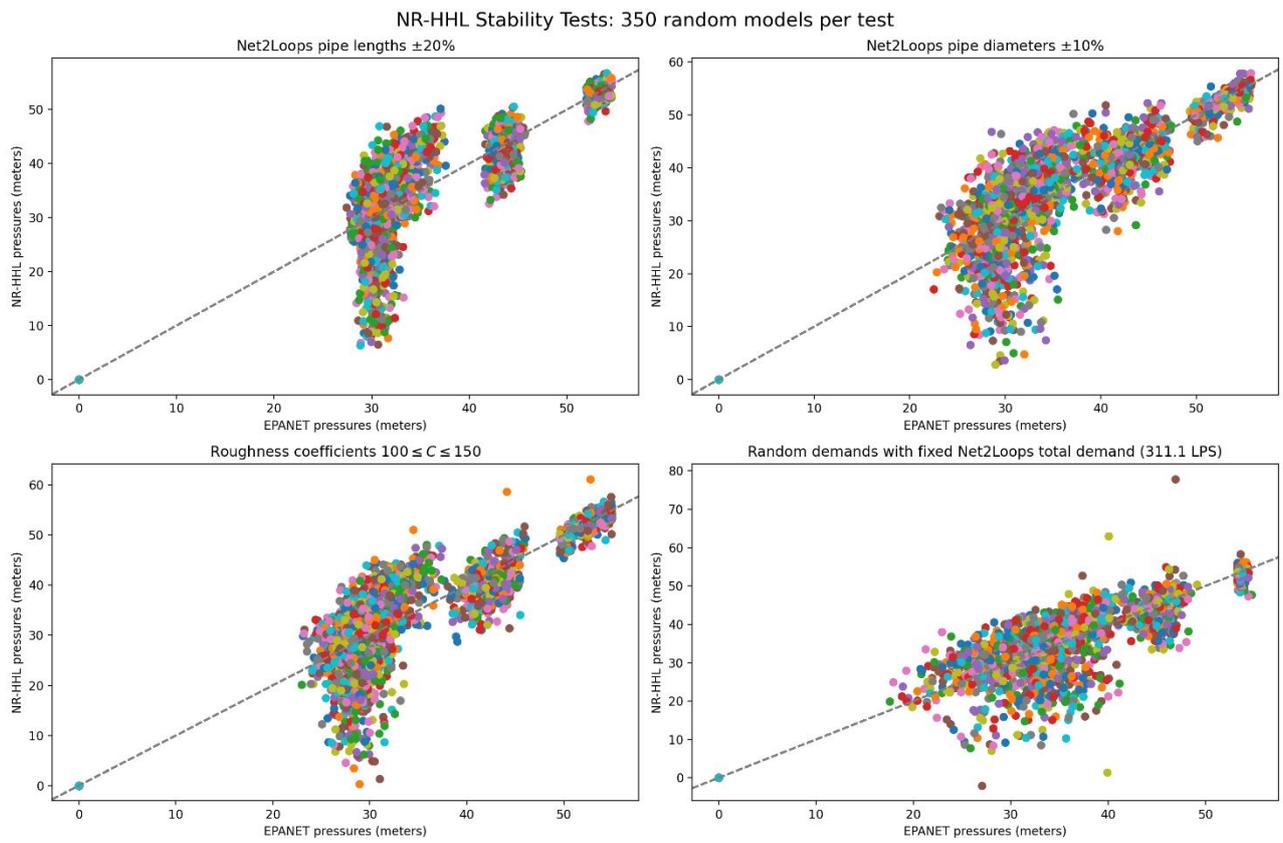

**Fig. 2.** Simulated water pressures using QNR-HHL for a 2-loop network for a wide range of pipe diameters, length, roughness and demand.

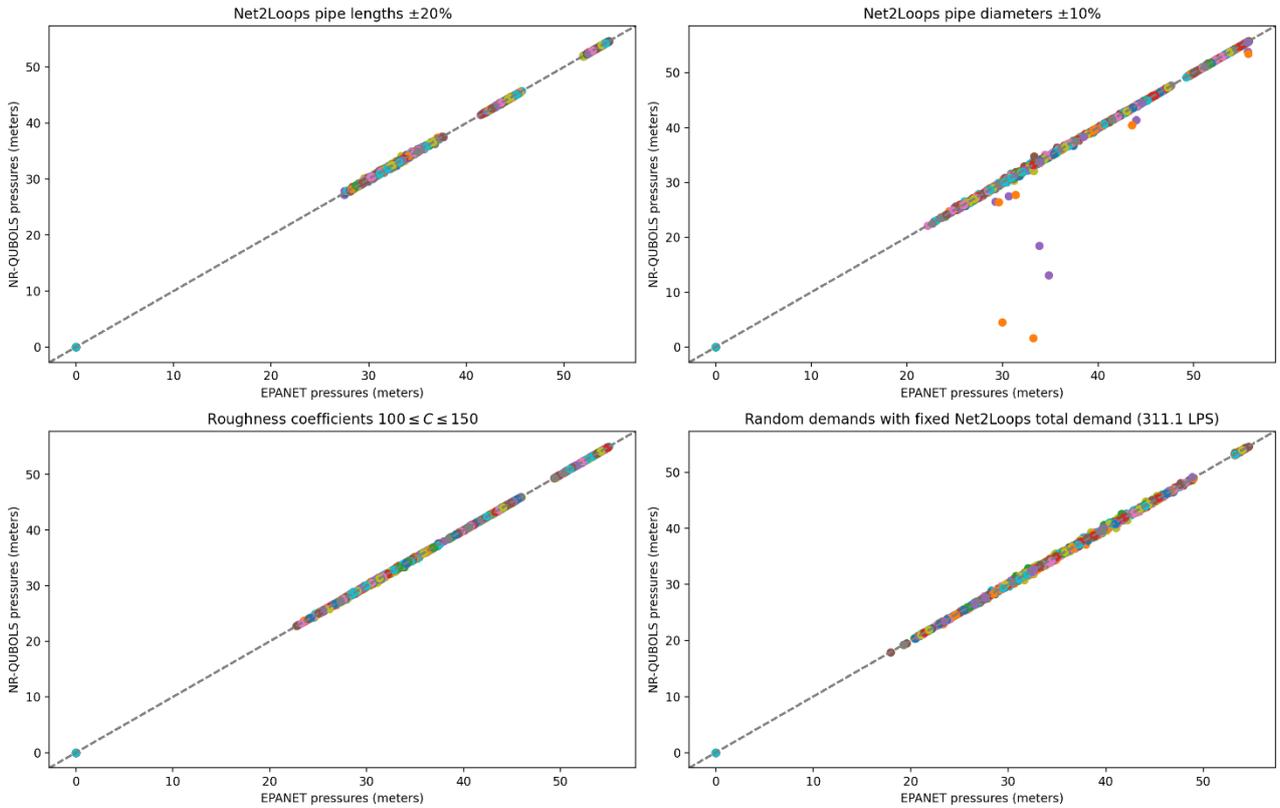

**Fig. 3.** Simulated water pressures using QNR-QUBOLS for a 2-loop network for a wide range of pipe diameters, length, roughness and demand.

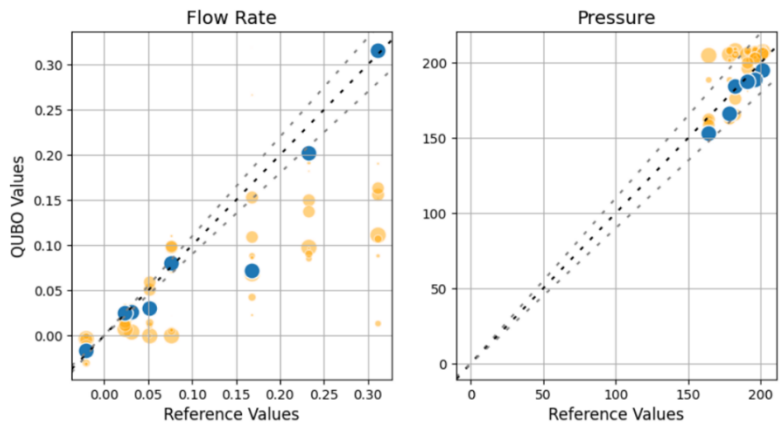

**Fig. 4.** Solution of the QUBO formulation of the hydraulics equations for the 0-loop network (top) and the 2-loop network (bottom). The minimum energy solution is marked in blue while the yellow markers indicate the results of other sampling paths.

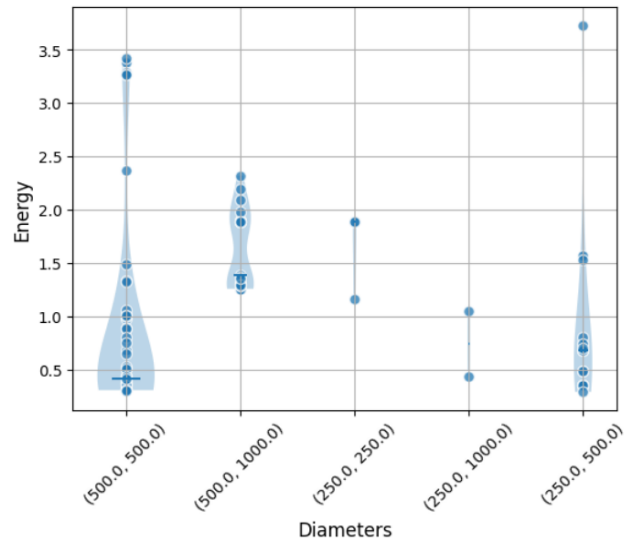

**Fig. 5.** Results of the optimization of the Least-Cost Design of the 0-loop network. The y-axis shows the energy of the cost function of the annealing algorithm, while the x-axis shows different combinations of pipe diameters in mm.

## 4. Conclusion

In conclusion, QNR-VQLS scored better than QNR-HHL and QNR-QUBOLS on simulating water flow and pressure for the 2-loop network. This is in line with expectations, as hybrid quantum-classical computing approaches are deemed more likely to lead to speed-ups on the near-term (Dalzell et al., 2023). To better understand the scalability of the QNR-VQLS solution, further research should focus on solving the hydraulics equations over larger systems.

QNR-HHL scored the worst out of the three versions of QNR in terms of accuracy. This could be explained by the problem of solving the gradient of a NR-GGA run over the hydraulics equations of WDN's being poorly conditioned in particular. HHL is more suitable for applications ran on large-scale fault-tolerant quantum computers further down the line.

The difference in accuracy between QNR-VQLS and QNR-QUBO is rather small on the tested dataset, where QNR-QUBO is plagued with a few outliers, most likely due to the QUBOLS subroutine having a few poor sampling runs. Results can potentially be improved by using more sampling runs in future works.

The higher-order QUBO method for simulating WDN's showed reasonable results but clearly started to become more prone to errors already for such a small network as the 2-loop network. This demonstrates that for the near future, potential advantages of using quantum computers will most likely be found in hybrid methods such as QNR. The application for solving LCD is technically feasible, but it was not possible to solve the problem over a larger network than the 0-loop network. This is likely due to the high overhead incurred by translating a high-order polynomial problem to a QUBO. Further work needs to be done to make claims regarding the scalability, such as running the algorithm on hybrid annealers and quantum annealers and improving the efficiency of the QUBO translation.

**Acknowledgements**

The Quantum Application Lab gratefully acknowledges financial support by the Quantum Delta NL Growth Fund program, as well as by the Municipality of Amsterdam through a SESA Grant. This work was submitted to IAHR 41[st] world congress on Innovative Engineering for Sustainable Development and is part of the conference proceedings.

**References**

Abbas A, King R, Huang HY, Huggins WJ, Movassagh R, Gilboa D, McClean J (2023) On quantum backpropagation, information reuse, and cheating measurement collapse, Advances in Neural Information Processing Systems, 36
Babayen A, Kapelan Z (2005) Least-cost design of water distribution networks under demand uncertainty, Journal of Water Resources Planning and Management, 131, 375-382
Bravo-Prieto C, LaRose R, Cerezo M, Subasi Y, Cincio L, Coles PJ (2023) Variational quantum linear solver, Quantum; The Open Journal for Science, 7, 1188pp
Chang CC, Gambhir A, Humble TS, Sota S (2019) Quantum annealing for systems of polynomial equations, Scientific Reports, 9
Dalzell AM, McArdle S, Berta M, Bienias P, Chen C, Gilyen A, Hann CT, Kastoryano MJ, Khabiboulline ET, Kubica A, Salton G, Wang S, Brandao FGSL (2023) Quantum algorithms: A survey of applications and end-to-end complexities, digital preprint, arXiv #2310.03011

Gouzien E, Sangouard N (2021) Factoring 2048-bit RSA integers in 177 days with 13 436 qubits and a multimode memory, Physical Review Letters, 127

Groenland K (2024) Timelines: when can we expect a useful quantum computer?, from *Introduction to Quantum Computing for Business*, chapter 4, Amsterdam University Press

Harrow AW, Hassidim A, Lloyd S (2009) Quantum algorithm for linear systems of equations, Physical Review Letters, 103

Klise K, Hart D, Bynum M, Hogge J, Haxton T, Murray R, Burkhardt J (2020) Water Network Tool for Resilience (WNTR) User Manual: Version 0.2.3. U.S. EPA Office of Research and Development, Washington, DC, EPA/600/R-20/185

Sonaje NP, Joshi MG (2015) A review of modelling and application of water distribution networks (WDN) softwares, International Journal of Technical Research and Applications, 3, 174-178

Rossman LA, Woo H, Tryby M, Shang F, Janke R, Haxton T (2020) EPANET 2.2 User Manual. U.S. Environmental Protection Agency, Washington, DC, EPA/600/R-20/133

Todini E, Pilati S (1998) A gradient algorithm for the analysis of pipe networks, Computer Applications in Water Supply, John Wiley & Sons, 1-20pp.

Todini E, Rossman LA (2012) Unified framework for deriving simultaneous equation algorithms for water distribution networks, Journal of Hydraulic Engineering, 139(5)

Yarkoni S, Raponi E, Bäck T, Schmitt S (2022) Quantum annealing for industry applications: introduction and review, Reports on Progress in Physics, 85